# Modeling the scaling properties of human mobility

Chaoming Song[1, 2*], Tal Koren [1, 2*], Pu Wang[1, 2], Albert-László Barabási[1, 2]

* These authors contributed equally to this work.

[1] *Center for Complex Network Research, Department of Physics, Biology and Computer Science, Northeastern University, Boston, MA 02115*

[2] *Department of Medicine, Harvard Medical School, and Center for Cancer Systems Biology, Dana Farber Cancer Institute, Boston, MA 02115*

**While the fat tailed jump size and the waiting time distributions characterizing individual human trajectories strongly suggest the relevance of the continuous time random walk (CTRW) models of human mobility, no one seriously believes that human traces are truly random. Given the importance of human mobility, from epidemic modeling to traffic prediction and urban planning, we need quantitative models that can account for the statistical characteristics of individual human trajectories. Here we use empirical data on human mobility, captured by mobile phone traces, to show that the predictions of the CTRW models are in systematic conflict with the empirical results. We introduce two principles that govern human trajectories, allowing us to build a statistically self-consistent microscopic model for individual human mobility. The model not only accounts for the empirically observed scaling laws but also allows us to analytically predict most of the pertinent scaling exponents.**

Uncovering the statistical patterns that characterize the trajectories humans follow during their daily activity is not only a major intellectual challenge, but also of importance for public health [1-5], city planning [6-8], traffic engineering [9, 10] and economic forecasting [11]. For example, quantifiable models of human mobility are indispensable for predicting the spread of biological pathogens [1-5] or mobile phone viruses [12].

In the past few years the availability of mobile phone records, GPS data, and other datasets capturing aspects of human mobility have given a new empirically driven momentum to the subject. While the available datasets significantly differ in their reach and resolution, the results appear to agree on a number of quantitative characteristics of human mobility. For example, both dollar bill tracking [13] and mobile phone data [14] indicate that the aggregated jump size ($\Delta r$) and waiting time ($\Delta t$) distributions characterizing human trajectories are fat-tailed, i.e. $P(\Delta r) \sim |\Delta r|^{-1-\alpha}$ and $P(\Delta t) \sim |\Delta t|^{-1-\beta}$ with $0 < \alpha \leq 2$ and $0 < \beta \leq 1$, where $\Delta r$ denotes the distances covered by an individual between consecutive sightings, and $\Delta t$ is the time spent by an individual at the same location. These findings suggest that human trajectories are best described as Lévy Flights (LF) or continuous time random walks (CTRW), a much studied modelling framework in the random walk (RW) community [13, 15-20].

The purpose of the present paper is to show, using a series of direct measurements, that human trajectories do follow several highly reproducible scaling laws. Yet, many of these laws are either not explained by the CTRW model, or they are in direct contradiction with the CTRW predictions, indicating the lack of modelling framework capable of capturing the basic features of human mobility. To explain the origin of the observed scaling laws, we introduce two principles that govern human mobility, serving as the starting point of a statistically acceptable microscopic model for individual human motion. We show that the model can account for the empirically observed scaling laws and allows us to analytically predict the pertinent scaling exponents.

**Scaling Anomalies** — We used two datasets to uncover the patterns characterizing individual mobility. The first dataset ($D_1$) captures for a one-year period the time-resolved trajectories of 3 million anonymized mobile phone users. Each time a user initiated or received a phone call the tower that routed the communication was recorded for billing purposes. Thus, the user's location was recorded with the resolution that is determined by the local tower density. The reception area of a tower varies from as little as a few hundred meters in metropolitan area to a few kilometres in rural regions, controlling our uncertainty about the user's precise location. However, since here we focus on the asymptotic scaling properties of human trajectories, these short distance uncertainties are not expected to affect our results (see Supplementary Material Section S1). The second dataset ($D_2$) uses the anonymized location record of 1,000 users who signed up for a location based service, thus their location was recorded every hour for a two week period. As a first step we calculated the displacement at hourly intervals, finding $P(\Delta r) \sim |\Delta r|^{-1-\alpha}$ with $\alpha = 0.55 \pm 0.05$ and an expected cutoff at $\Delta r \sim 100$ km, corresponding to the distance people could reasonably cover in an hour. We used the $D_2$ dataset to measure $P(\Delta t)$, where the waiting time $\Delta t$ is defined as the time a user spent at one location. We find that $P(\Delta t)$ follows $P(\Delta t) \sim |\Delta t|^{-1-\beta}$ with $\beta = 0.8 \pm 0.1$ and a cutoff of $\Delta t = 17$ hours, likely capturing the typical awake period of an

individual. Taken together, the fat-tailed nature of $P(\Delta r)$ and $P(\Delta t)$ suggest that human follow a CTRW during their daily mobility. Next we discuss three empirical observations that indicate that human trajectories follow reproducible scaling laws, but also illustrate the shortcoming of the CTRW model in capturing the observed scaling properties:

(A) *The number of distinct locations* $S(t)$ visited by a randomly moving object is expected to follow [21-23]

$$S(t) \sim t^{\mu}, \qquad (1)$$

where $\mu = 1$ for Lévy flights [24] and $\mu = \beta$ for CTRW. Interestingly, our measurements indicate that for humans $\mu = 0.6 \pm 0.02$ (see Fig. 1a), smaller than the CTRW prediction of $\beta = 0.8 \pm 0.1$. The fact that $\mu < 1$ indicates a slow-down at large time scales, a deceasing tendency of the user to visit previously unvisited locations.

(B) *Visitation frequency*: The probability $f$ of a user to visit a given location is expected to be asymptotically ($t \rightarrow \infty$) uniform everywhere ($f \sim$ const.) for both LF and CTRW. In contrast, the visitation patterns of humans is rather uneven, so that the frequency $f$ of the $k^{\text{th}}$ most visited location follows Zipf's law [14]

$$f_k \sim k^{-\zeta}, \qquad (2)$$

where $\zeta \approx 1.2 \pm 0.1$ (see Fig. 1b). This suggests that the visitation frequency distribution follows $P(f) \sim f^{-(1+1/\zeta)}$.

(C) *Ultra-slow diffusion*: The CTRW model predicts that the mean square displacement (MSD) asymptotically follows $\langle \Delta x^2(t) \rangle \sim t^{\nu}$ with $\nu = 2\beta/\alpha \approx 3.1$. Since both $P(\Delta r)$ and $P(\Delta t)$ have cutoffs, asymptotically the MSD should converge to a Brownian behaviour with $\nu = 1$. However, this convergence is too slow[25] to be relevant in our observational time frame. Either way, CTRW predicts that the longer we follow a human trajectory, the further it will drift from its initial position. Yet, humans have a tendency to return home on daily basis, suggesting that simple diffusive processes, that are not recurrent in two dimensions, do not offer a suitable description of human mobility. Indeed, our measurements indicate an ultra-slow diffusive process, in which the MSD appears to follow a slower than logarithmic growth (see Fig. 1c and Ref. 14). Such ultra-slow growth of the MSD is rare in diffusion, having been observed before only in a few disordered systems, from glasses (for example the Sinai model [26]) to polymers [27] and iterated maps [28].

On one end, the findings summarized in A - C indicate that individual human mobility does follow reproducible scaling laws, whose origins remain to be uncovered. Yet, they also document systematic deviations from the predictions of the LF or CTRW based null models.

The main purpose of this paper is to offer a model that not only explains the origin of the anomalies A - C, but also leads to a self-consistent statistical model of individual human mobility.

**Generic Mechanisms and Individual Mobility Model** — As we build our model, we will take for granted the observations that the jump size $P(\Delta r)$ and the waiting time $P(\Delta t)$ distributions characterizing individual human trajectories are heavy tailed, a phenomenon addressed by a series of models [29-33]. Yet, $P(\Delta r)$ and $P(\Delta t)$ alone are not sufficient to explain the scaling laws A - C. We propose that the main reason for this discrepancy is that two generic mechanisms, exploration and preferential return, both unique to human mobility, are missing from the traditional random walk (LF or CTRW) models:

(1) *Exploration*: Random walk models assume that the next diffusive step is independent of the previously visited locations. In contrast, the scaling law (1) indicates that the tendency to explore additional locations decreases with time. Indeed, the longer we observe a person's trajectory, the harder is to find locations in the vicinity of their home/workplace that they have not yet visited.

(2) *Preferential Return*: In contrast with the RW based models for which the visitation probability is random and uniform in space, humans show significant propensity to return to the locations they visited frequently before, like home or workplace.

In what follows we present an individual mobility (IM) model that incorporates ingredients (1) and (2), showing that they are sufficient to explain the anomalies A - C. The model, intended to describe the trajectory of an individual, assumes that at time $t = 0$ the individual is at some preferred location (see Fig. 2). After a waiting time $\Delta t$ chosen from the $P(\Delta t)$ distribution, the individual will change its location. We assume that the individual has two choices:

(i) *Exploration*: With probability

$$P_{\text{new}} = \rho S^{-\gamma} \tag{3}$$

the individual moves to a new location (different from the $S$ locations it visited before). The distance $\Delta r$ that it covers during this exploratory jump is chosen from the $P(\Delta r)$ distribution and its direction is selected to be random. As the individual moves to this new position, the number of previously visited locations increases from $S$ to $S$+1.

(ii) *Preferential Return*: With the complementary probability $P_{\text{ret}} = 1 - \rho S^{-\gamma}$ the individual returns to one of the $S$ previously visited locations. In this case, the probability $\Pi_i$ to visit location $i$ is chosen to be proportional to the number of visits the user previously had to that location. That is, we assume that

$$\Pi_i = f_i, \tag{4}$$

an assumption known as preferential attachment or cumulative advantage in network and social science [34-38].

**Model Predictions** — The IM model has two parameters, $0 < \rho \leq 1$ and $\gamma \geq 0$, both of which control the user's tendency to explore a new location during its next move vs. returning to a previously visited location. The numerical values of these two parameters will be determined later from the empirical data.

*Anomaly A:* To explain the origin of the anomaly A, we note that in the IM model the probability that an individual moves to a new location is proportional to $S^{-\gamma}$, i.e. $dS/dn \propto S^{-\gamma}$, predicting $S \sim n^{1/(1+\gamma)}$, where $n$ is the total number of discrete moves the individual had up to time $t$. For a fat-tailed waiting time distribution $P(\Delta t) \sim |\Delta t|^{-1-\beta}$ the time $t$ scales with the number of jumps $n$ as $t \sim n^{1/\beta}$ (Section S3A in Supplementary Material), obtaining that $S(t)$ follows (1) with exponent

$$\mu = \beta/(1+\gamma). \tag{5}$$

To verify the validity of this prediction for the IM model, in Fig. 1d we calculated $S(t)$ numerically for different values of $\beta$, finding that the asymptotic scaling exponent of $S(t)$ follows Eq. (5). Therefore, we predict that $\mu \leq \beta$, in line with the empirical data.

*Anomaly B:* To account for anomaly B we notice that $m_i$, the number of visits to location $i$, increases as $dm_i/dn = \Pi_i(1-P_{\text{new}})$, where $\Pi_i = f_i = m_i/\Sigma_i m_i(n)$ is the probability to return to the location $i$ during step $n$. When $\gamma > 0$, in the limit of $S(t) \to \infty$ the probability to explore a new location is negligible compared to the return visits, thus asymptotically we have $dm_i/dn = m_i/\Sigma_i m_i(n)$. Since $\Sigma_i\, m_i(n) = n$, we obtain $m_i(n) = n/n_i$, where $n_i$ denotes the jump during which location $i$ was first visited, at which moment $m_i(n_i) = 1$. Thanks to preferential return (4), the earlier a location is visited, the more it is visited later on. Thus the ranking $k_i$ for location $i$ coincides with the order in which it was first visited, i.e. $k_i = S(n_i) \sim n_i^{1/(1+\gamma)}$. Since the visitation frequency $f_i$ is proportional to $m_i(n) = n/n_i$, we have $f_k \sim k^{-\zeta}$ with exponent $\zeta = 1 + \gamma$. In general, we find (Section S3C in Supplementary Material)

$$\zeta = \begin{cases} 1+\gamma, & \gamma > 0 \\ 1-\rho, & \gamma = 0 \end{cases}. \tag{6}$$

Note that for $\gamma = 0$ and $\rho = 1$ Eq. (6) predicts $\zeta = 0$, indicating, as expected, that the visitation becomes homogeneous in the CTRW limit.

To test the validity of prediction (6), in Fig. 1e we measured $f_k$ for different values of $\gamma$ and $\rho$ in the IM model, finding that the numerically observed scaling behaviour is in agreement with Eq. (6). The model also suggests that the observed Zipf's law is rooted mainly in the preferential return (4). Indeed, if we calculate $f_k$ for a "democratic" model in which $\Pi_i$ is independent of $f_i$ (i.e. the individual visits the previously visited locations with the same likelihood), Zipf's law vanishes (see Fig. 1e).

Anomaly C: To understand the origin of anomaly C, we note that the number of jumps to new locations $l$ relates to the displacement $\Delta r$ as $\Delta r \sim l^{1/\alpha}$ (Section S3A in Supplementary Material), suggesting that $\langle \Delta r^2 \rangle \sim \langle l^{2/\alpha} \rangle = \sum_{l=1}^{n} l^{2/\alpha} P(l \mid S)$, where $P(l|S)$ is the probability that the $S^{\text{th}}$ location is $l$ steps away from the starting point of the individual. Note that $l$ is different from $n$, as $l$ counts only moves that result in a jump to some new location, so $l \le n$. $P(l|S)$ follows the recurrent equation $P(l \mid S) = \sum_{i=1}^{S} P(l-1 \mid k) f_k^S$ , where each term within the sum represents a jump from the $k^{\text{th}}$ location to $S^{\text{th}}$ location. Here $f_i^s$ is the probability of visitation of the $k^{\text{th}}$ location given that the total number of locations visited previously is $S$-1, and is well approximated by Zipf's law (2) $f_k^s \approx (\zeta\text{-}1)/(1\text{-}S^{1-\zeta})\ k^{-\zeta}$. By applying this approximation (see Section S3D in Supplementary Material), we find

$$\langle \Delta x^2 \rangle^{\alpha/2} \sim \log\left(\frac{1-S^{1-\zeta}}{\zeta-1}\right) + const, \qquad (7)$$

which relates the MSD to the number of distinct location $S$ visited by the user. In Figs. 1c &1f, we plot the MSD vs. $S$ for different values of $\zeta$, finding that Eq. (7) agrees well with the numerical results. Furthermore, in the inset of Fig 1c we plot the prediction (7) against the empirical data, again with excellent agreement.

To determine the explicit time dependence of MSD, we use Eqs. (1) and (5) to predict three possible scaling regimes for large $t$ (or $S$):

(a) For $\zeta < 1$ ($\gamma = 0$ and $\rho < 1$), $S^{1-\zeta}$ diverges and thus the MSD grows with time as $(\log t)^{2/\alpha}$. In this regime the visitation frequency $P(f) \sim f^{(1+1/\zeta)}$ has exponent greater than two and thus the diffusion is dominated by the infrequently visited locations.

(b) For $\zeta = 1$, we have $(\zeta\text{-}1)/(1\text{-}S^{1-\zeta}) = \log(S)$, and thus the MSD $\sim \log(\log(t))^{2/\alpha}$.

(c) For $\zeta > 1$, $S^{1-\zeta}$ approaches zero for large $S$ and thus the MSD is expected to saturate. This is because $P(f)$ decays with an exponent whose value is less than two and thus the individual's motion is dominated by his/her most visited location.

To compare further the IM model with the empirical data, we first need to test the validity of the hypotheses (i) - (ii) as formulated by Eqs. (3) and (4). To do this, we used the $D_2$ dataset and measured the rate at which the users visit new locations (Fig. 3a). We find that $P_{\text{new}} \propto S^{-\gamma}$ with $\gamma = 0.21 \pm 0.02$, a result that not only confirms the validity of hypothesis (3), but also provides the numerical value of the exponent $\gamma$. To test the validity of the hypothesis (ii), in Fig. 3b we plot the probability $\Pi$ that a user returns to a previous location in function of the previous visitation frequency $f$ of this location. The plot indicates that $\Pi = f$, confirming the validity of preferential return (4). Furthermore, the ratio between $P_{\text{new}}$ and $S^{\gamma}$ measures the parameter $\rho$ for each user, allowing us to plot $P(\rho)$ for our user group, finding that $P(\rho)$ approximately follows a normal distribution with the mean at $<\rho> \approx 0.6$. Note that the scaling properties of the IM model do not depend on $\rho$ for $\gamma > 0$, thus while the precise value of $\rho$ is important to parameterize the model, it does not affect the scaling laws discussed in A - C.

Finally, to match our model based prediction with the empirical data, it is necessary to inspect the relationship between the exponents characterizing the model and those observed for real human mobility. In this respect our starting point are the three independently determined exponents $\alpha = 0.55 \pm 0.05$, $\beta = 0.8 \pm 0.1$ and $\gamma = 0.21 \pm 0.02$. Equation (5) predicts $\mu = 0.67 \pm 0.07$, in agreement within the error bar with the empirical value $\mu = 0.6 \pm 0.02$ (Fig 1a). Furthermore, Eq. (6) predicts $\zeta = 1.21 \pm 0.02$, again in excellent agreement with the empirical value $\zeta = 1.2 \pm 0.1$. Moreover, Eq. (5) and Eq. (6) predict the existence of the scaling relationship $\beta = \mu\zeta$ between three empirical exponents (valid for $\zeta > 1$), which is again consistent with empirical data (Fig. 1b). Finally, when it comes to anomaly C the analytical predictions offer three different scaling regimes, determined by the value of $\zeta$. The empirical data indicates the Zipf exponent $\zeta = 1.2 \pm 0.1$, for which we predict a saturation in the MSD. To understand how the system reaches saturation, we expand $\zeta$ around $\zeta = 1$, finding that in the transient regime the MSD scales as $(\log\log(t))^{2/\alpha}$. This prediction is valid only if $S(t) < \exp(1/(\zeta-1)) \approx 148$, which is true for 89% of the users that have $S < 148$ over the one year period. This prediction is consistent with the empirical data shown in Fig. 1c, documenting a slower than logarithmic growth of the MSD. Note that in the empirical data we do not observe the predicted saturation of the MSD, potentially due to the finite time frame (1 year) used in the study. To estimate the saturation time, we extrapolated Eq. (1) up to $S(t_s) = 148$ for the empirical data, predicting that the saturation can be reached only after $t_s = 5$ years, beyond our data horizon. Note that the MSD saturation could also be rooted in the finite size of the country (i.e. the finite number of towers a user can visit). Yet, scaling arguments (see Supplementary Material S6) indicates that the saturation time for $S$ due to the finite size effects is approximately 10,000 years. Therefore, the saturation of MSD is rooted in the need to return to the most visited locations, forcing the new locations explored by a user to be close to user's most visited location.

Another key empirical observation is population heterogeneity: the radius of gyration $r_g$ of the trajectory of different individuals is found to follow a fat tailed distribution (see Fig. 4a and Ref. 14). As we show in Fig. 4c, we find that our model can reproduce this feature as well, indicating that the fat tailed $P(r_g)$ is a consequence of the inherent fluctuations present within the model and it is rooted in the $P(\Delta r)$ distribution. Indeed, in Supplementary Section S4 we show that the tail of $P(r_g)$ and $P(\Delta r)$ are expected to share same exponent $1+\alpha$. Furthermore, we find that the model reproduces not only the $P(r_g)$ distribution, but the ultra-slow growth of $r_g$ as well (Fig. 4d), in agreement with the empirical data (Fig. 4b).

It is important to note that in contrast with the traditional RW, LF or CTRW models, our model is dynamically quenched. That is, after an individual explores a new location, he/she will have an increasing tendency to return to it in the future, generating a recurrent and relatively stable mobility pattern for each individual. In principle one could also consider a model that assigns a quenched visitation variable to each site. Our approach not only avoids the need to parameterize such a model, but also achieves the trajectory selection dynamically, through its self-quenching character.

We also note that the model is designed to capture the long term spatial and temporal scaling patterns, thus in its present from it does not reproduce the short time frame temporal order and correlations potentially present in individual mobility. Our choice to focus the asymptotic properties is driven not only by theoretical arguments (we aim to reproduce the universal and not the transient patterns), but also by practical considerations: many human mobility driven processes, from epidemic spreading to city planning, are driven by the asymptotic characteristics of human mobility. To achieve a better short range temporal fidelity, we need to incorporate the periodic modulations that are known to characterize human mobility (there is a 24-hour and 7 days periodicity in human mobility; individuals are less likely to change locations during night and quite mobile in the morning and late afternoon, see Supplementary Material S7) as well potential correlations in spatial mobility (i.e. if location B is between locations A and C in space, the likely order of visitation will be A→B→C or C→B→A). These correlations further constrain the human trajectories, being partly responsible for the high degree of the predictability characterizing individual mobility patterns [39]. Finally, we note that the dynamical quenching and the recurrent behaviour is unique to human trajectories, and does not restrict banknote diffusion, or foraging behaviour [13, 33, 40]. As such, our model represents an improvement over the CTRW/LF models, as it is adapted to capture the specifics of human mobility. By reproducing the basic scaling laws characterizing human trajectories, the current model offers a conceptual framework that has the flexibility to absorb future extensions, potentially improving the temporal fidelity of its short term dynamics as well.

**Methods** — *Mean Square Displacement (MSD)*: Due to the significant population heterogeneity of $P(r_g)$, MSD averaged over all users diverges and thus is ill-defined. We therefore grouped users based on their $r_g$, guaranteeing that the MSDs defined within groups do not diverge. We calibrate $t = 0$ to correspond to the moment when an individual leaves his/her most visited location. In order to compare with the analytical prediction (7), the displacement is counted only when the individual explores a new site. This methodology does not affect the scaling behaviour of regular random walk based models.

*Number of Distinct Locations (S)*: Similar to the MSD, $S(t)$ is the average over different $r_g$ groups. Given the large population in our dataset, the standard error of mean is quite small and the error bar in Fig.1a is covered by the symbols. The error bar in $\mu$ mainly comes from fitting of the power law, whose correlation coefficient $r = 0.998$ and the $p$-value associated with the fit is less than $10^{-6}$.

**Acknowledgements:** We thank Marta Gonzales, Dashun Wang, James Bagrow and Zehui Qu for discussions and comments on the manuscript. This work was supported by the James S. McDonnell Foundation 21st Century Initiative in Studying Complex Systems; NSF within the Information Technology Research (DMR-0426737), and IIS-0513650 programs; the Defense Threat Reduction Agency Award HDTRA1-08-1-0027 and the Network Science Collaborative Technology Alliance sponsored by the U.S. Army Research Laboratory under Agreement Number W911NF-09-2-0053.

**Author contributions**: C.S., T.K. and A-L.B. conceived and executed the research; C.S and T.K ran the numerical simulations; C.S, T.K. and P.W. analysed the empirical data.

**Correspondence to**: Albert-László Barabási[1, 2] e-mail: alb@neu.edu

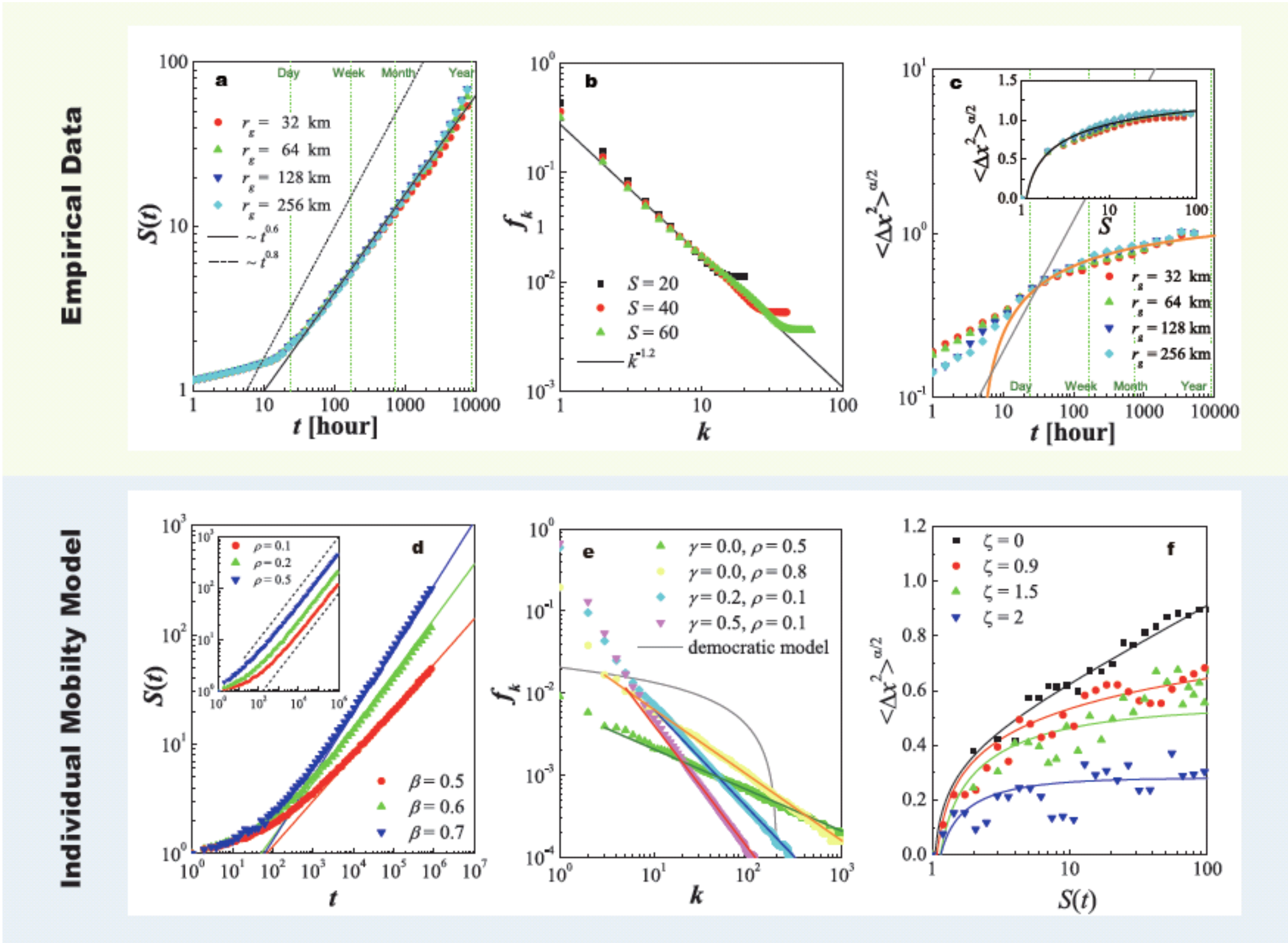

Fig. 1: **Empirical results vs. the predictions of the individual mobility (IM) model.** (a) The number of visited distinct locations $S(t)$ vs. time for different $r_g$ groups, indicating that $S(t)$ grows as $t^{\mu}$, with $\mu \approx 0.6 \pm 0.02$ (straight line). The dashed line represents the prediction of CTRW $S(t) \sim t^{\beta}$ with $\beta \approx 0.8$. (b) Zipf's plot showing the visitation frequency $f_k$ of the $k$-th most visited location of a user for different $S$ values. The empirical data is well approximated by $f_k \sim k^{-\xi}$, where $\xi \approx 1.2 \pm 0.1$. (c) Time evolution of the MSD ($\langle \Delta x^2 \rangle^{\alpha/2}$) in log-log scale for user groups with different radius of gyration $r_g$, where the MSDs are normalized by their value at $t$ = 1 year. The orange curve represents the analytical prediction for the asymptotic behavior. The grey line represents the analytical prediction of CTRW, $\langle \Delta x^2 \rangle^{\alpha/2} \sim t^{\beta}$ with $\beta$ = 0.8. Inset: the normalized MSD vs. $S$ for different $r_g$ groups, where the black curve represents the analytical prediction. (d) The number of visited distinct locations $S(t)$ vs. time in a log-log plot, as predicted by the IM model with different $\beta$ values ($\alpha = 0.5$, $\gamma = 0.2$ and $\rho = 0.1$). The straight lines represent the analytical prediction. Inset: the IM model prediction for different $\rho$ values ($\alpha = 0.5$, $\beta = 0.6$ and $\gamma = 0.2$. (e) Zipf's plot showing the visitation frequency $f_k$ in the IM model with different $\gamma$ and $\rho$ values ($\alpha = 0.5$, $\beta = 0.6$). The straight lines show the analytical prediction. The grey symbols correspond to a democratic model in which preferential attachment is absent. (f) the MSD ($\langle \Delta x^2 \rangle^{\alpha/2}$) vs. the number of visited distinct locations $S(t)$ in a log-linear scale. The symbols correspond to simulations with different $\xi$ values, where $\xi = 0$ corresponds to the democratic model. The continuous lines represent the analytical prediction.

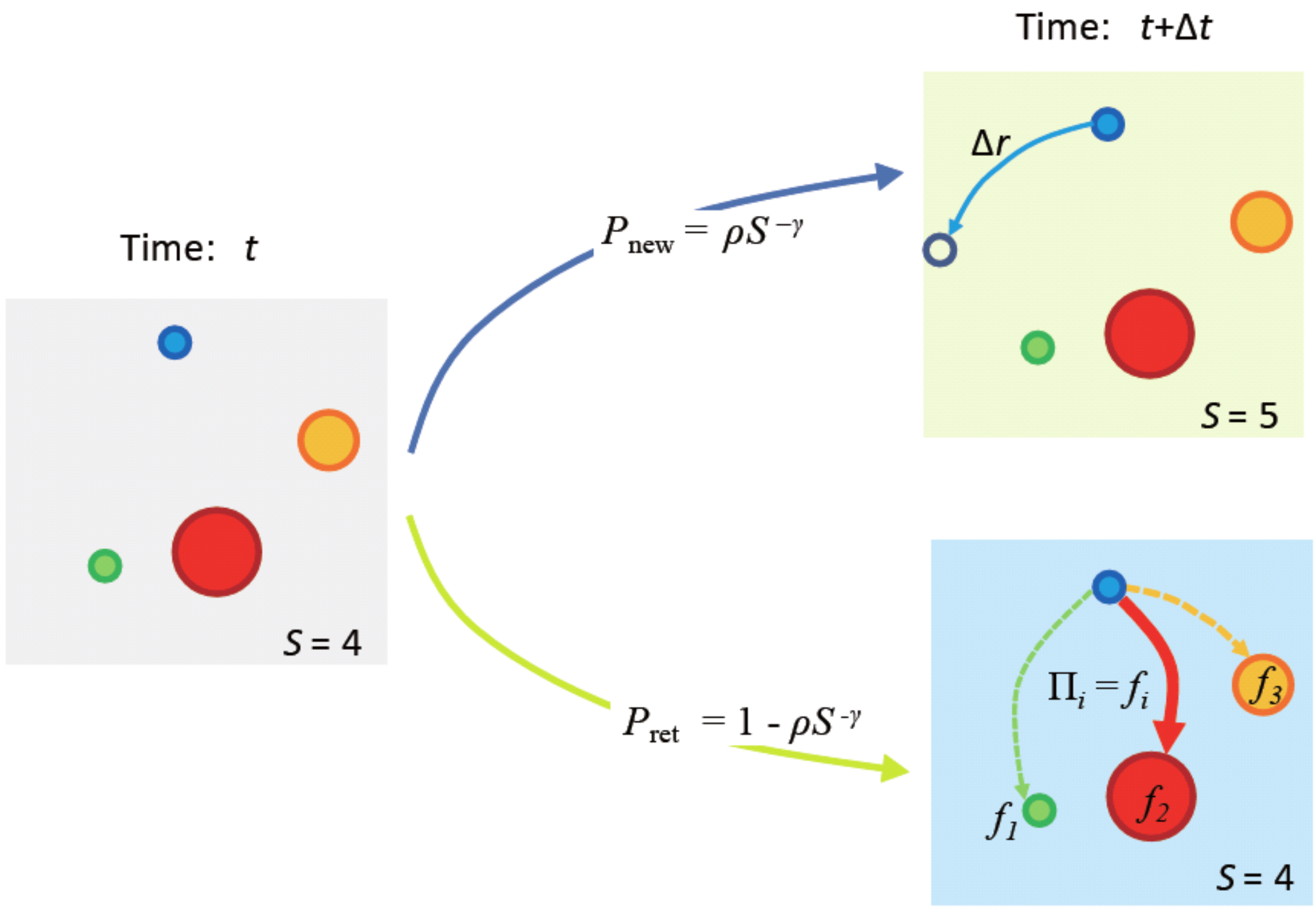

Fig. 2: **Schematic description of the Individual Mobility (IM) model.** Starting at time $t$ from the configuration shown in the left panel, indicating that the user visited previously $S$ = 4 locations with frequency $f_i$ that is proportional to the size of circles drawn at each location, at time $t + \Delta t$ (with $\Delta t$ drawn from the $P(\Delta t)$ fat tailed distribution) the user can either (i) Exploration (upper panel): visit a new location at distance $\Delta r$ from its current location, where $\Delta r$ is chosen from $P(\Delta r)$ fat tailed distribution, or (ii) Preferential return (lower panel): return to a previously visited location with probability $P_{ret} = 1 - \rho S^{-\gamma}$, where the next location will be chosen with probability $\Pi_i = f_i$.

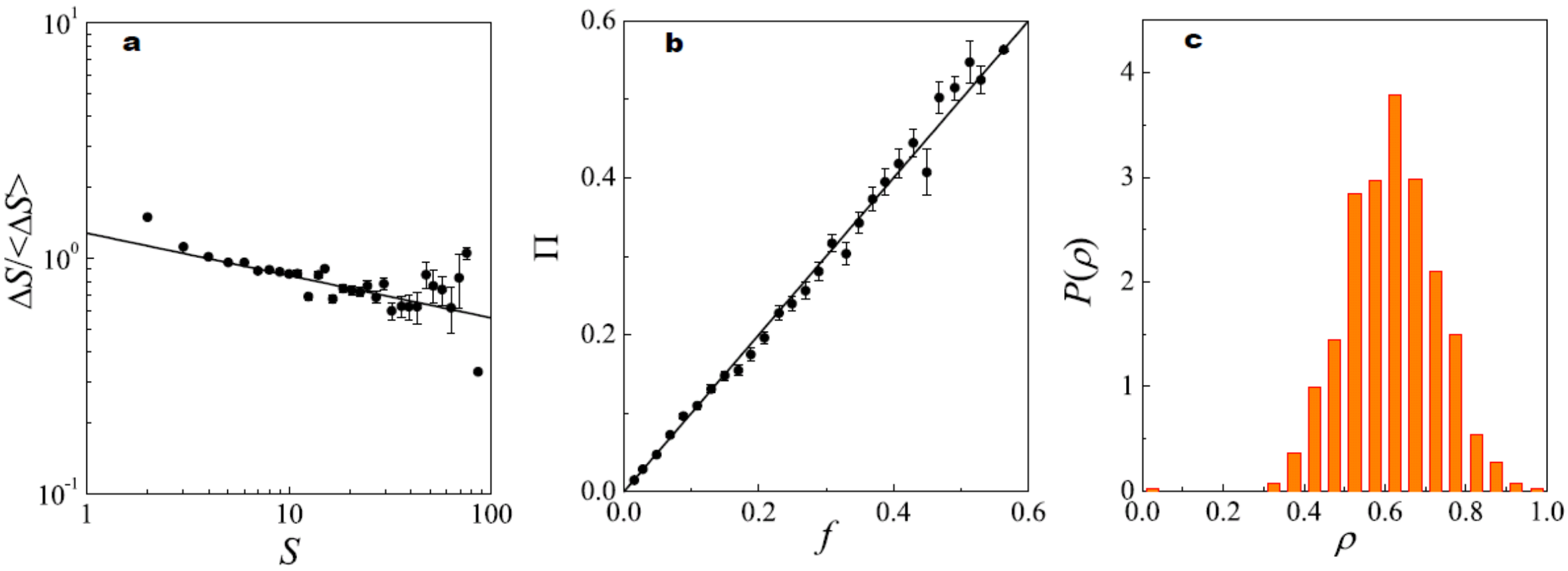

Fig 3: **Testing the hypotheses behind the IM model.** (a) The changes in the number of distinct locations $\Delta S/\langle \Delta S\rangle$ visited by the users vs. the number of distinct locations $S$ visited previously, indicating that $P_{\text{new}} \sim \Delta S \sim S^{-\gamma}$ with $\gamma$ = 0.21±0.02. The error bars correspond to standard error of the mean (SEM). (b) The probability to return to a previously visited location Π in function of the previous frequency of visitation $f$, offering evidence for preferential attachment $\Pi = f$. The error bars associate with SEM. (c) The probability density function (PDF) $P(\rho)$ for the user group, finding that $P(\rho)$, where $\rho$ is the model parameter (see Fig. 2), approximately follows a normal distribution with mean $\rho \approx 0.6$.

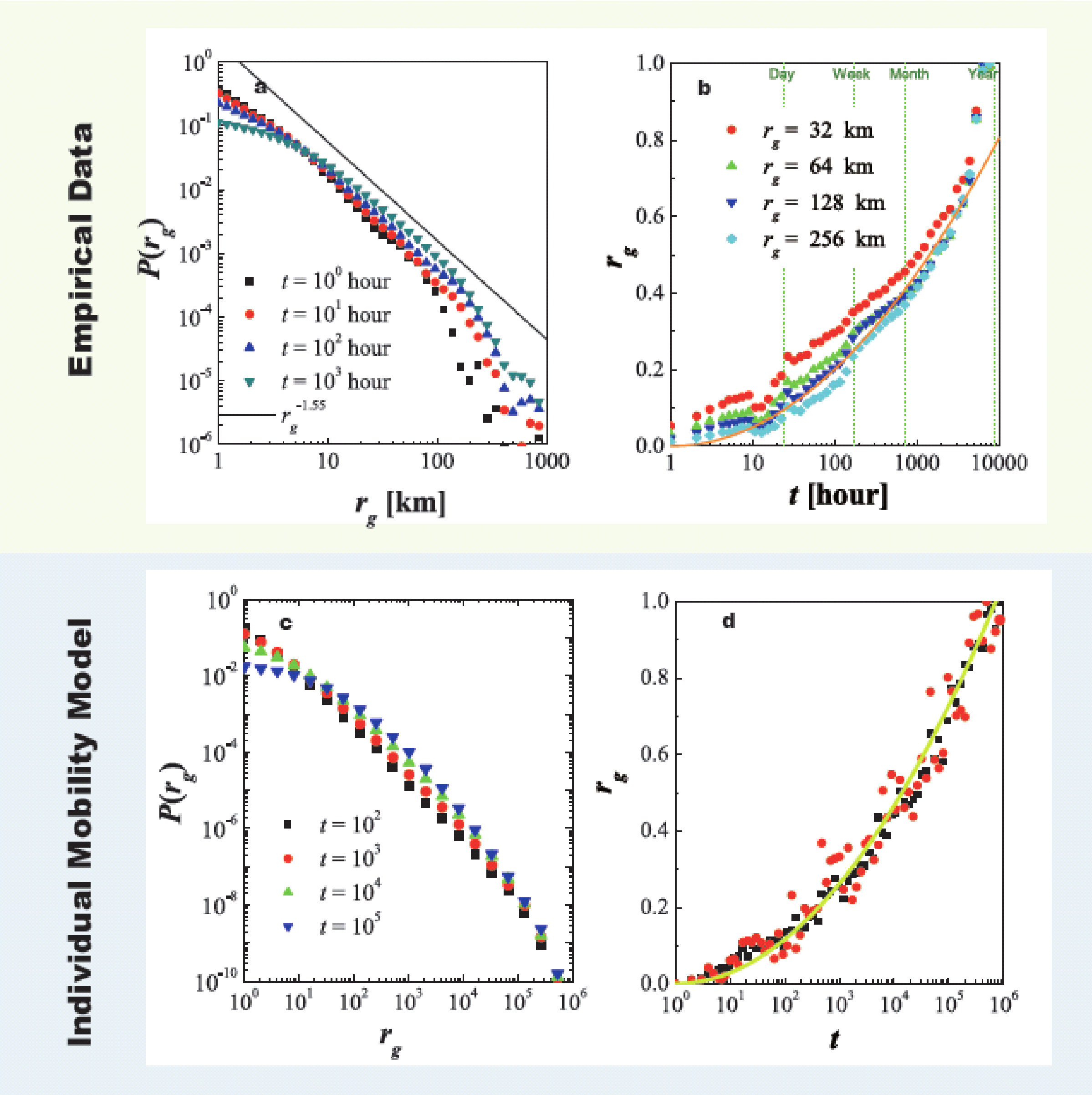

Fig. 4: (a) The $P(r_g)$ distribution of the radius of gyration $r_g$ for the mobile phone users at different moments of time. The straight line, shown as a guide to the eye, represents a power law decay with exponent $1+\alpha \approx 1.55$. (b) The time evolution of radius of gyration $r_g(t)$ for the mobile phone users, where $r_g(t)$ is normalized by its maximal value $r_g(T)$ at time $T = 1$ year. The orange curve is a guide to the eye, following $(\log t)^2$. (c) The distribution of radius of gyration $P(r_g)$ for the IM model using $\alpha = 0.75$, $\beta = 0.6$, $\gamma = 0.2$ and $\rho = 0.1$, the values found to be of direct relevance to human mobility. (d) The time evolution of radius of gyration $r_g(t)$ for the IM model with the same parameters as (c), where $r_g(t)$ is normalized by its value $r_g(T)$ at $T = 10^6$. The green curve is a guide to the eye, following $(\log t)^2$.